# Electrically-driven control of nanoscale chemical changes in amorphous complex oxide memristive devices

Wilson Román Acevedo[1], Myriam H. Aguirre[2,3,4], Diego Rubi[1,*]

[1]Instituto de Nanociencia y Nanotecnología (INN),
CONICET-CNEA, Gral. Paz 1499, 1650 San Martín, Argentina

[2] Instituto de Nanociencia y Materiales de Aragón (INMA-CSIC), Campus Río Ebro C/Mariano Esquillor s/n, 50018 Zaragoza, Spain

[3] Dpto. de Física de la Materia Condensada, Universidad de Zaragoza, Pedro Cerbuna 12, 50009 Zaragoza, Spain

[4] Laboratorio de Microscopías Avanzadas, Edificio I+D, Campus Rio Ebro C/Mariano Esquillor s/n, 50018 Zaragoza, Spain

# Abstract

Our study demonstrates that strong cationic segregation can occur in amorphous complex oxide memristors during electrical operation. With the help of analytic techniques, we observed that switching the electrical stimulation from voltage to current significantly prevents structural changes and cation segregation at the nanoscale, improving also the device cycle-to-cycle variability. These findings could contribute to the design of more reliable oxide-based memristors and underscore the crucial effect that has type of electrical stimulation applied to the devices has on their integrity and reliability.

## I. INTRODUCTION

Memristive systems are capacitor-like micro or nanostructures able to change their electrical resistance -usually in a non-volatile way- upon the application of electrical stimulation [1]. Initially, the technological importance of memristors was mostly linked to the possibility of developing Resistive Random Access Memories [2]; however, in the past few years, their ability to mimic the (analog) behavior of biological synapses triggered a great deal of research as they are thought to constitute one of the key building blocks for the development of neuromorphic computing [3], intended to outperform standard Von-Neumann computers in data intensive tasks. At the moment, several proof-of-concept neuromorphic devices based on memristors have been developed and tested [4–6].

Memristive mechanisms usually rely on the reversible electromigration at the nanoscale of charged defects, leading to resistance changes in the device. In the particular case of insulating oxides, these defects are usually charged oxygen vacancies [7], which can form conducting nanofilaments that bridge both metallic electrodes [8]. On the other hand, oxygen vacancy dynamics can induce changes at metal/oxide interface resistance via Schottky-barrier height modulation [1] or redox-reaction [9], leading to an area-distributed memristive effect.

Despite the mechanism of electrochemical metalization, where metallic ions coming from Ag or Cu electrodes form a conducting filament [10, 11], the possibility of chemical changes linked to cation electromigration has been less studied in oxide-based memristors [12, 13]. These changes, which can be self-accelerated by local Joule heating and should strongly depend on the specific protocol of external stimulation, might produce drastic effects in the composition, nanostructure and, therefore, the memristive performance of oxide-based devices. The obtention of local information at the nanoscale [14] is essential to achieve a deep understanding and, thefore, control of the memristive response of these devices.

* diego.rubi@gmail.com

It was previously shown in manganite-based devices that the memristive response strongly depends on the type of stimulation; for instance, the use of voltage stimulation produced a two-level memristive response with high cycle-to-cycle variations, while the use of current stimulation allowed the stabilization of a third resistance level [15, 16], together with improved cycle-to-cycle variation and endurance. This difference can be rationalized in terms of the limited power injection during the SET process (high to low resistance transition) for the case of current stimulation, that limits self-heating effects and the concomitant nanostructural and chemical changes. For current stimulation, the electrical power P is given by $P = I^2 R$, where I and R are the current and the resistance, respectively. In this case, the electrical power remains self-limited for the SET transition as I is externally controlled and R switches to a state of lower resistance. For the case of voltage stimulation, the electrical power is given by $P = V^2/R$, with V being the applied voltage. The SET transition is now linked to a power peak, related to the drop in R while V remains externally fixed. We notice that the uncontrolled power peak exists even if a compliance current is used, as electronic sources usually have a limited time response that can't avoid an unwanted power overshoot in a time-window of the order of ms, necessary to stabilize the compliance current.

In the present paper we study the chemical changes at the nanoscale on devices based on amorphous thin films of the complex oxide $La_{0.5}Sr_{0.5}Mn_{0.5}Co_{0.5}O_3$ (LSMCO) under different electrical stimulation. This system presents, in its perovskite crystalline form, the existence of both oxidized an reduced phases, with different conductivities, linked via a topotactic redox transition [17], which can be triggered electrically. When sandwiched be- tween $Nb:SrTiO_3$ (n-type material) and Pt electrodes, crystalline LSMCO (p-type) presents a memristive effect concomitant with a strong memcapacitive one [18, 19]. The origin of this behavior relies on the formation of a switchable n-p diode, where the oxidation and reduction of LSMCO strongly changes the balance between n- and p-carriers at the interface, which controls the depletion layer and thus the interface capacitance and the electrical leakage. It was shown that the multi-mem behavior can be improved by tuning the out-of-plane orientation, the type of electrical stimulation [19] and by incorporating dopants such as Ce [20], which enhance the perovskite reducibility by producing a steric effect that enlarges the perovskite unit cell. A similar enhancement of the reducibility was shown by controlling the unit cell deformation using substrates with different lattice mismatch with the LSMCO unit cell [21].

Here we address, instead, LSMCO films grown at low, CMOS-compatible, temperatures, leading to amorphous films. In this case, cations are not placed in well defined minima of the crystalline structure and therefore are more prone to migrate during memristive operation. Our results show that the use of voltage or current as electrical stimulation has a dramatic effect on the structure and chemistry at the nanoscale of the devices. In the latter case, chemical and nanostructural changes are minimized and this leads to a more stable memristive behavior, with a lower cycle-to-cycle variability. These results might contribute to the design of memristive devices with higher reliability and better suited to endure electrical operation.

## II. EXPERIMENTAL

LSMCO thin films were grown on NSTO single crystals by Pulsed Laser Deposition, using a Nd:YAG laser ($\lambda$ = 266 nm). The deposition temperature was set in 200 °C and the oxygen background pressure was 0.085 mbar. The film thickness, as determined from x-ray reflectivity (Figure S1), was $\approx$ 38.5 nm. X-ray Bragg-Brentano scans (not shown here) display the only presence of reflections arising from the substrate, confirming the amorphous nature of the films. Top Pt electrodes ($\approx$ 29 nm thick) were microfabricated by standard UV litography and magnetron sputtering. The device area was $\approx 2.4 \times 10^4 \ \mu m^2$. The bottom NSTO electrode was grounded and electrical bias was applied to the top Pt electrode, by using a Keithley 2612 Source Meter Unit (SMU) hooked to a commercial probe-station. Memristive characterization consisted in dynamic current-voltage (IV) curves, where voltage (current) pulses of electrical stimulation, with varying amplitude and polarity according to the scheme depicted in the inset of Figure 1(b), were applied and current (voltage) was measured during the application of the pulses. In addition, we measured on the same run the so-called Hysteresis Switching Loops (HLS) [22, 23], which allows tracking the evolution of the remnant resistance immediately after the application of each write pulse. For that, a small reading voltage pulse (typically 100-200 mV) is applied in between write pulses, and the remnant resistance is extracted by measuring the current. High resolution Scanning Transmission Electron Microscopy (HRTEM) with High Angular Annular Dark Field Detector (STEM-HAADF) was performed with a FEI Titan G2 microscope with probe corrector (60-300 keV). Local chemical analysis was performed by Energy Dispersive Spectroscopy (EDS) and Electron Energy Loss Spectroscopy (EELS).

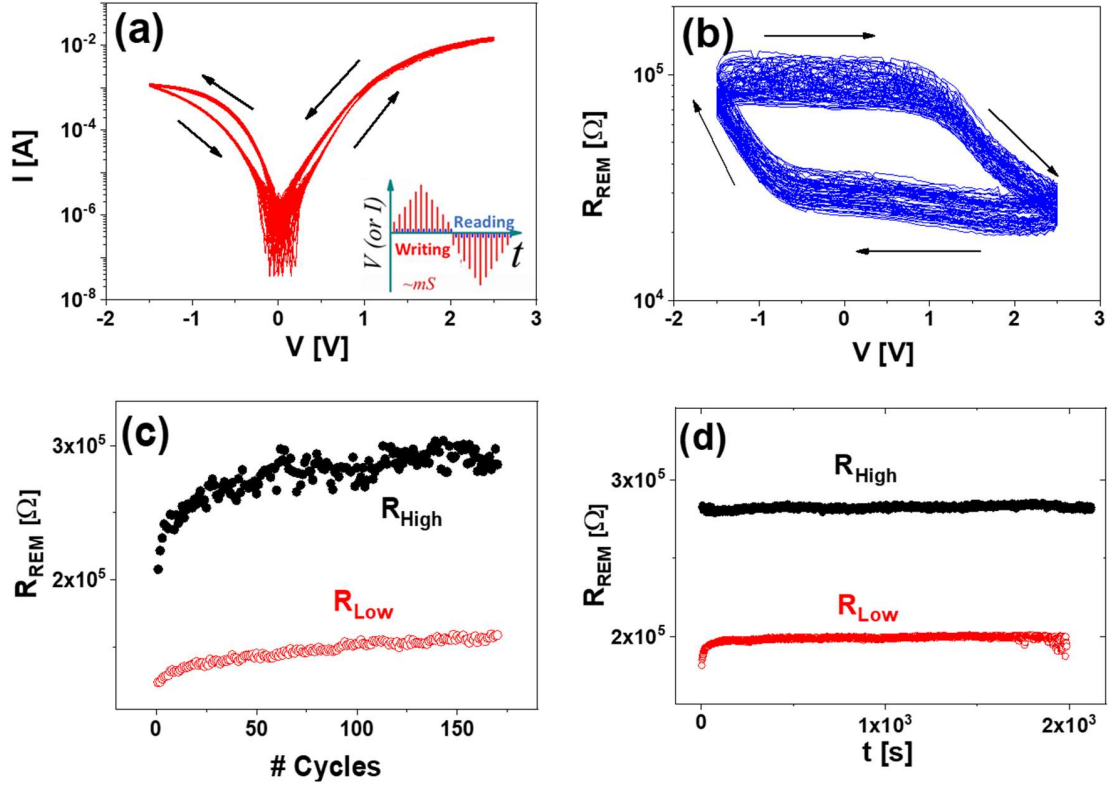

FIG. 1: (a) Dynamic pulsed I–V curve measured on a NSTO/LSMCO/Pt device stimulated with voltage. The arrows indicate the chirality of the curve; (b) Hysteresis switching loop recorded on the same device. The inset in panel (a) sketches the protocol of electrical stimulation; (c) Cycle-to-cycle variation test performed by applying single SET and RESET voltage pulses; (d) Retention experiments for both $R_{LOW}$ and $R_{HIGH}$ states.

The devices were electroformed by applying single pulses of either voltage or current, which produced a drop of the resistance from the virgin state of $\approx 10^5$ Ω to $\approx 80$ Ω and $\approx 200$ Ω, respectively, as shown in Figure S2.

## III. RESULTS

Figures 1(a) and (b) show dynamic IV curves and HSLs, respectively, recorded on a device stimulated with voltage during 100 cycles. The presence of hysteresis in both cases is a signature of the memristive effect. The SET transition is observed for ≈ 2.5 V, while the inverse transition (low to high resistance, RESET) takes place for ≈ -1.5 V. The high and low resistance levels extracted from the HSL are $R_{HIGH}$ ≈ $1x10^5$ Ω and $R_{LOW}$ ≈ $2x10^4$ Ω, respectively, giving an ON-OFF ratio of ≈ 5. Figure 1(c) shows cycle-to-cycle variation experiments, recorded by applying single pulses of 2.5 V and -1.5 V to switch the device between $R_{HIGH}$ and $R_{LOW}$, for 170 cycles. It is found that both resistance states present a drift towards higher resistance values -which is more pronounced for $R_{HIGH}$ and the first ≈ $10^2$ s after the start of the test- upon cycling-. We also notice that the ON-OFF ratio in this case is lower than that obtained from the HLS. This is related to the fact that the resistance states were set by applying single pulses, on the contrary of the HSL where both $R_{HIGH}$ and $R_{LOW}$ are achieved from a pulsed ramp that produces a cumulative effect on the resistance changes. Finally, Figure 1(d) displays retention time experiments for both $R_{HIGH}$ and $R_{LOW}$ up to $2x10^3$ s.

Figures 2(a) and (b) show the measured IV curve and HSL on the device stimulated with current. The SET (RESET) transition is observed for $5x10^{-3}$ A ($-2x10^{-3}$ A). Resistance levels are $R_{HIGH}$ ≈ $10^5$ Ω and $R_{LOW}$ ≈ $10^4$ Ω, respectively, with an ON-OFF ratio of ≈ 10, about twice the value obtained for the voltage-controlled case. Figure 2(c) displays cycle-to-cycle variation tests recorded after applying single pulses of $5x10^{-3}$ A and $-2x10^{-3}$ A to trigger the SET and RESET transitions, respectively. In this case, the drift of both $R_{HIGH}$ and $R_{LOW}$ upon pulsing is milder that in the case of voltage stimulation. Again, the ON-OFF ratio obtained for the cycle-to-cycle variation test is smaller than the one obtained from the HSL due to the different voltage pulsing scheme used to reach $R_{HIGH}$ and $R_{LOW}$ in each case. Figure 2(d) display retention times up to $2x10^3$ s.

Further information about the (different) cycle-to-cycle response obtained for both type of stimulation can be seen from Figures 3(a) and (b), which show the calculated cumulative probabilities (F) for both $R_{HIGH}$ and $R_{LOW}$ and voltage and current stimulation, respectively. The inspection of this Figure evidences that the $R_{HIGH}$ and $R_{LOW}$ curves corresponding to the current-controlled device are steeper and comprise narrower resistance

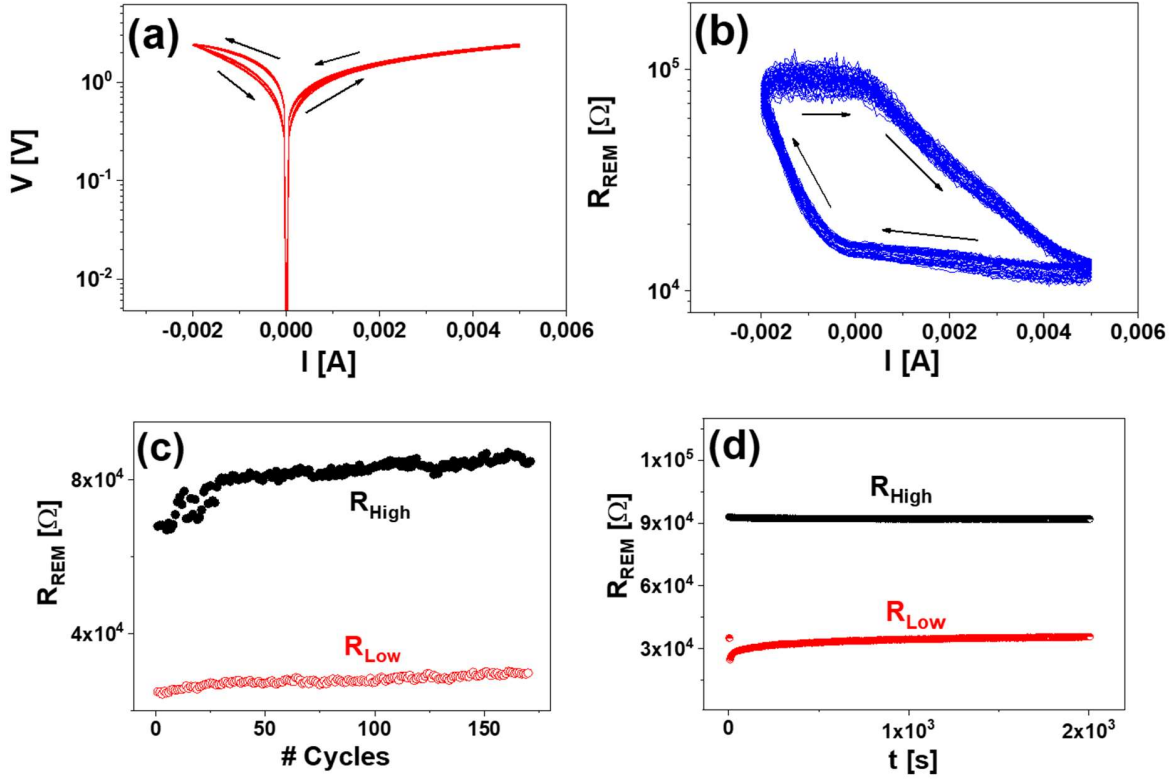

FIG. 2: (a) Dynamic pulsed I–V curve recorded on a NSTO/LSMCO/Pt device stimulated with current. The arrows indicate the chirality of the curve.; (b) Hysteresis switching loop recorded simultaneously on the same device; (c) Cycle-to-cycle variation test measured on the same device; (d) Retention experiments recorded for both $R_{LOW}$ and $R_{LOW}$ states

windows, reflecting a lower cycle-to-cycle variability.

It has been shown [24] that the evolution of both $R_{HIGH}$ and $R_{LOW}$ states upon repeated cycling can be described by a Weibull probability distribution f given by

$$f(R;\lambda;k) = \frac{k}{\lambda}\left(\frac{R}{\lambda}\right)^{k-1} e^{-\left(\frac{R}{\lambda}\right)^k} \qquad (1)$$

where $\lambda$ is the scale parameter, k is the shape parameter and R is the resistance. This gives the following expression for the cumulative probability F

$$F(R;\lambda;k) = 1 - e^{-(R/\lambda)^k} \qquad (2)$$

which can be linearized as follows

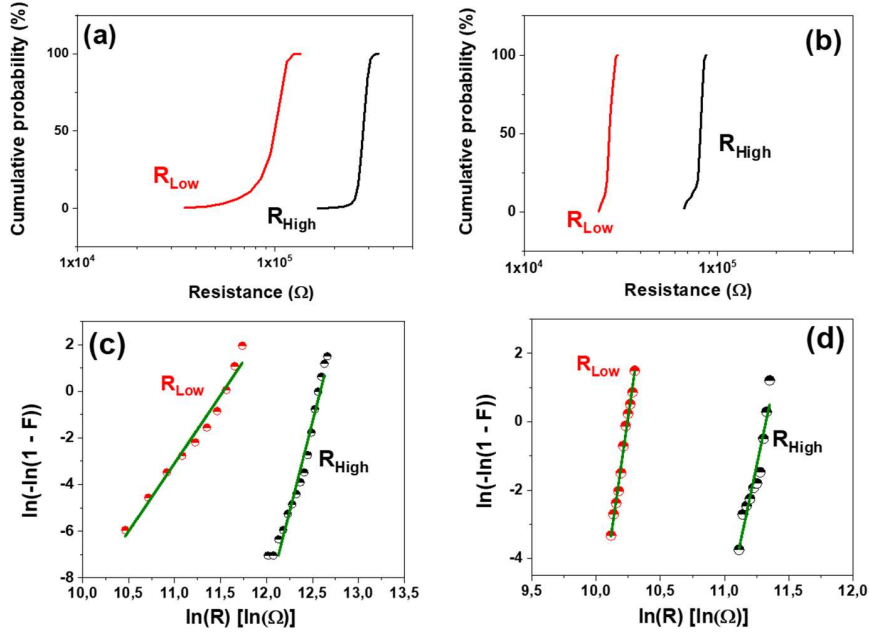

FIG. 3: Cumulative probabilities, obtained from cycle-to-cycle variation tests, for $R_{HIGH}$ and $R_{LOW}$ states and voltage (panel (a)) and current (panel (b)) stimulation; (c), (d) Weibull fittings for $R_{HIGH}$ and $R_{LOW}$ states and voltage and current stimulation, respectively.

$$\ln(-\ln(1 - F(R; \lambda, k))) = k \ln R - k \ln \lambda \qquad (3)$$

Using the experimental values obtained for F, in Figures 3(c) and (d) we have plotted $ln(-\ln(1 - F))$ as a function of $ln(R)$, being found a linear behavior -with slopes given by the parameter k- for both $R_{HIGH}$ and $R_{LOW}$, validating the assumption of cycle-to-cycle variations described by Weibull probability distributions. By performing linear fittings we have extracted values of $\lambda$ and k for both types of stimulation. For voltage stimulated devices, we obtained, for $R_{HIGH}$, $\lambda = 2.9 \times 10^5$ $\Omega$ and k = 15.4 and, for $R_{LOW}$, $\lambda = 1.0 \times 10^5$ $\Omega$ and k = 5.8. For current operated devices we obtained, for $R_{HIGH}$, $\lambda = 8.2 \times 10^4$ $\Omega$ and k = 17.7 and, for $R_{LOW}$, $\lambda = 2.8 \times 10^4$ $\Omega$ and k = 26.0. The lower cycle-to-cycle variations of the current stimulated device is reflected in the higher values obtained for k for both resistive states, related to narrower Weibull distributions (the limit case k → ∞ corresponds to a Dirac delta function with no cycle-to-cycle variations).

We recall that device-to-device variations might be highly relevant in memristive devices and need to be taken into account during electrical characterization [25]. Figure S3 shows Weibull plots calculated from cycle-to-cycle variation experiments performed on different devices controlled either with voltage and current, confirming that the already mentioned trend - that is current controlled devices present lower cycle-to-cycle variations with narrower Weibull distributions- is seen for all the tested devices.

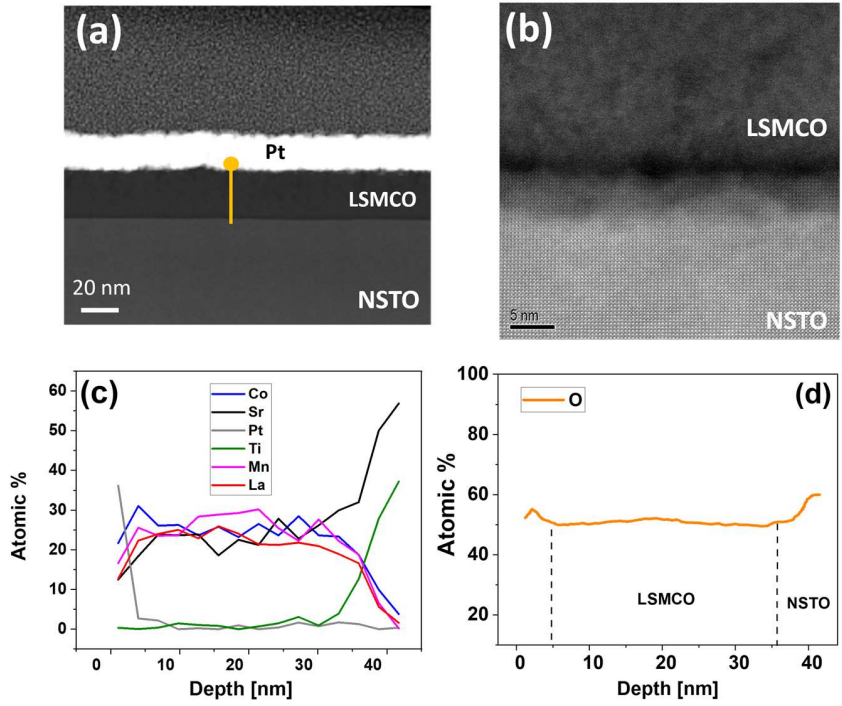

FIG. 4: (a) Low-magnification STEM-HAADF cross-section corresponding to a virgin NSTO/LSMCO/Pt device; (b) High-resolution STEM-HAADF cross section recorded on the same device; (c) EDS line scans quantifying Sr, Ti, Mn, La, and Co cations. The scan, indicated in (a) with a yellow line, goes from the LSMCO/Pt interface to the NSTO substrate; (d) EELS oxygen line scan recorded on the same sample.

It can thus be concluded that current stimulation seems to produce a more stable memristive behavior. In order to get a better understanding at the nanoscale of the scenario linked to this behavior, we have performed HRTEM experiments on virgin devices, together with ones stimulated with both voltage and current.

Figure 4(a) shows a low magnification STEM-HAADF image corresponding to a virgin device. LSMCO and Pt thicknesses of ≈ 38.5 nm and ≈ 29.0 nm are seen. The high reso-lution image of Figure 4(b) shows the crystalline character of the NSTO substrate together with the amorphous nature of the LSMCO layer. This is confirmed by the Fast Fourier Transforms displayed in Figure S4. Figure 4(c) shows an EDS line-scan that quantifies the cationic composition. It is observed a LSMCO cationic stochiometry consistent with the nominal one (25 % at. for La, Sr, Mn and Co). The oxygen composition was quantified from the EELS line-scan displayed in Figure 4(d), evidencing some oxygen deficiency ($\approx$ 50 % at.) in relation to the nominal one (60 % at.).

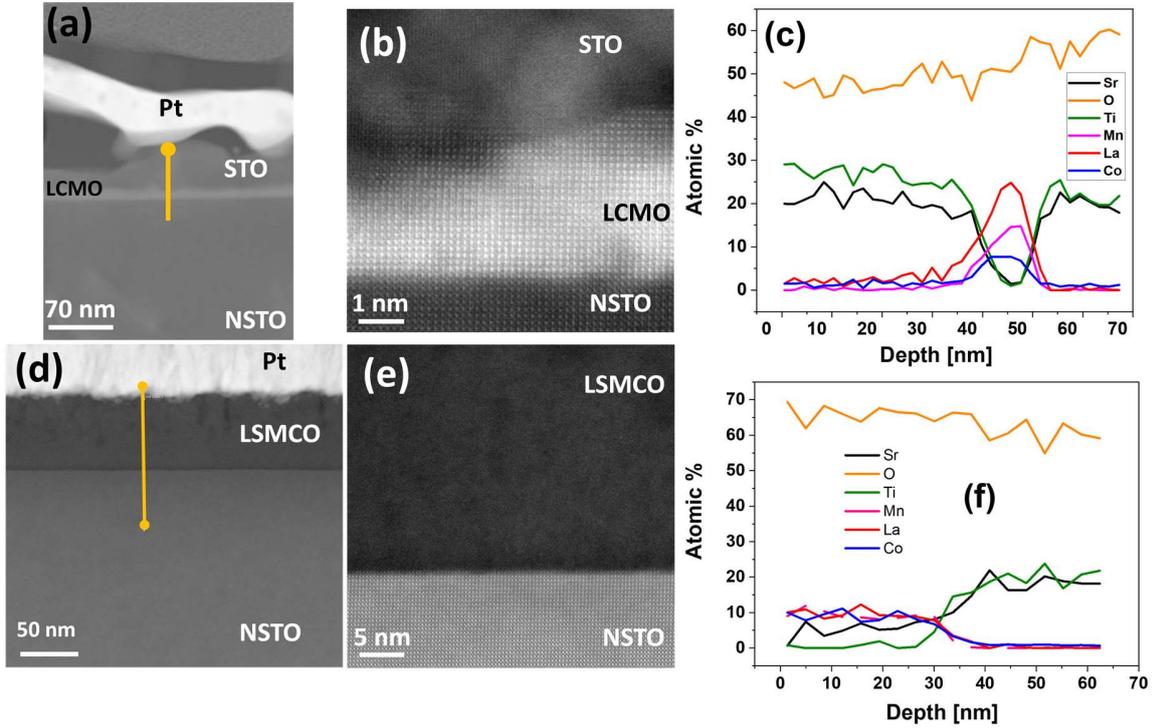

FIG. 5: (a), (d) Low magnification STEM-HAADF cross-sections recorded on NSTO/LSMCO/Pt devices stimulated with voltage and current, respectively; (b), (e) High-resolution STEM-HAADF cross-sections measured on the same devices; (c), (f) EDS line scans quantifying Sr, Ti, Mn, La, Co, and O elements in NSTO/LSMCO/Pt devices stimulated with voltage and current, respectively. As indicated with yellow lines in (a) and (d), the scans start at the LSMCO/Pt interface and end in the NSTO substrate.

Figure 5(a) shows a low magnification STEM-HAADF cross-section corresponding to a device stressed with voltage. Several changes can be seen; the most important one is related to the observation that the LSMCO layer is splitted in two layers with dissimilar HAADF contrast, indicating that they have different chemical composition. Figure 5(b) shows a high magnification STEM-HAADF image, displaying that the bottom (brighter) layer is crystalline, while the top one (darker) seems to remain amorphous. This is confirmed by the Fast Fourier Transforms shown in Figure S5. In order to obtain further information about the chemistry of these layers, we have performed an EDS line-scan, which is displayed in Figure 5(c). It is seen that the top layer consists in *SrTiO$_3$* (STO) while the bottom one presents a chemical composition consistent with the one of *La$_2$CoMnO$_6$* (LCMO) perovskite. We notice that both STO/LCMO and LCMO/NSTO interfaces are not completely sharp and there is some cationic intermixing, which is more pronounced in the former case. The described results evidence that the application of voltage stimulation triggers a strong cationic diffusion that modifies the LSMCO chemistry, probably accelerated by self-heating effects that also induces the (partial) crystallization of the oxide layer [26]. Figure 5(a) also shows the existence of a meandered top Pt electrode and the presence of "voids" at the Pt/oxide interface, also reflecting the effect of the high temperatures achieved during device operation [27].

Figure 5(d) shows a low magnification STEM-HAADF cross-section corresponding to a device stimulated with current. Despite the presence of some extended defects, appearing with darker contrast and located closer to the top interface with the Pt electrode, the LSMCO layer displays a rather uniform STEM-HAADF brightness, suggesting the absence of cation segregation, unlike the case of the device operated with voltage (recall Figures 5(a)- (c)). This is confirmed by the high magnification cross-section of Figure 5(e) and by the EDS line-scan displayed in Figure 5(f), which shows that the nominal cation stochiometry is maintained in the device operated with current, without significant differences with regard the virgin device. The oxygen content, instead, seems to be slightly higher in this device, suggesting some oxygen uptake in the LSMCO layer during the device operation, likely via diffusion through the Pt top electrode grain boundaries [28]. Figure 5(e) also suggests that the amorphous character of the LSMCO layer is preserved upon stimulation with current.

## IV.  DISCUSSION AND CONCLUSSIONS

The presented data indicate substantial differences in the nanostructure and chemistry of the devices operated either with voltage or current.  For the former case, the partial crystallization of the oxide layer, together with strong cation diffusion that changes the chemistry of the layer, is likely related to the high temperatures achieved during device operation for this type of electrical stimulation, and is linked to a higher cycle-to-cycle variability. The memristive mechanism can be related to the electromigration of oxygen vacancies between *SrTiO$_3$* and *La$_2$CoMnO$_6$* layers, which are both n-type insulators that locally increase their conductivities in the presence of oxygen vacancies [29, 30]. For instance, a positive voltage applied to the top electrode triggers the movement of oxygen vacancies from *SrTiO$_3$* to *La$_2$CoMnO$_6$*, leading to a local increase of *SrTiO$_3$* layer resistance and a decrease of the *La$_2$CoMnO$_6$* one. The fact that an overall SET transition is observed for the device indicates that the latter effect dominates the resistance drop.

On the other hand, the device operated with current displayed no significant crystallization process, together with a maintained chemical composition in relation to the pristine device. Lower cycle-to-cycle variations were observed in relation to the voltage operated device. We relate this observations with the self-limited electrical power injection in current controlled mode. The memristive mechanism in this case is likely related to the electromigration of oxygen vacancies between the LSMCO layer (p-type) and the NSTO substrate (n-type), presenting an interface with a rectifying behavior, as shown in Figure S6 (i.e. an n-p diode is formed). Assuming that the number of donors at NSTO ($N_D \approx 5 \times 10^{18} cm^{-3}$ [31]) is much lower than the number of acceptors at the LSMCO layer ($N_A \approx 10^{22} cm^{-3}$, arising from the $\approx 0.5$ holes/f.u. related to the substitution of half of the $La^{3+}$ ions by $Sr^{2+}$), the diode depletion layer W should depend on $N_D$ as $W \sim N^{-0.5}$ [32]. The injection of oxygen vacancies in the NSTO crystal (achieved with positive voltage) increases $N_D$ and therefore reduces W, lowering in this way the interface resistance.

In summary, our results show that strong cationic segregation can take place in amorphous memristors during electrical operation. This is likely related to the lack of long-range atomic ordering and with the existence of reduced energy barriers for ion movement, together with the absence of extended defects, such as dislocations, that usually hinder ion diffusion. Our results show that the switching of the electrical stimulation from voltage to current allows to significantly reduce cationic segregation, helping to maintain the device structure and chemistry at the nanoscale and improving the memristive performance by reducing cycle- to-cycle variations. These results could help to the design of more reliable oxide-based memristors, pointing also out the key importance of the type of electrical stimulation on the memristive performance.

**Supplementary Material**

See Supplementary Material for additional x-ray reflectivity and transmission electron microscopy experiments and electrical measurements.